\documentclass[twocolumn]{openjournal}

\usepackage[english]{babel}
\usepackage{graphicx}
\usepackage{natbib}
\usepackage{placeins}
\usepackage{amsmath}
\usepackage{float}
\usepackage[dvipsnames]{xcolor}

\newcommand{\Mopcs}{\mathrm{M}_\odot\,\mathrm{pc}^{-2}}
\newcommand{\kms}{\mathrm{km}\,\mathrm{s}^{-1}}

\newcommand{\dex}{\mathrm{dex}}

\providecommand{\ion}[2]{#1\,\textsc{#2}}
\renewcommand{\ion}[2]{#1\,\textsc{#2}}

\shorttitle{Inferring total halo mass using photometry}
\shortauthors{Brook et al.}

\begin{document}

\title
{\bfseries Inferring the total mass of galaxy halos using photometry} 
\author{Chris Brook\altaffilmark{1,2}}
\author{Arianna di Cintio\altaffilmark{1,2}}
\author{Claudio Dalla Vecchia\altaffilmark{2,1}}
\author{Salvador Cardona Barrero\altaffilmark{1,2}}
\author{Elena Arjona-G\'{a}lvez\altaffilmark{1,2.3}}

\affil{$^1$ Universidad de La Laguna, E-38200 La Laguna, Tenerife, Spain}
\affil{$^2$ Instituto de Astrof\'isica de Canarias, C/ V\'ia L\'actea s/n, E-38205 La Laguna, Tenerife, Spain}
\affil{$^3$ Max Planck Institute for Astronomy, Königstuhl 17, D-69117 Heidelberg, Germany
}

\begin{abstract}
Galaxy size, as defined as the radius at which the stellar surface density reaches $1~\Mopcs$ ($R_1$), has been shown to correlate more tightly with stellar mass than the effective radius. Cosmological simulations further predict a tight relation between $R_1$ and dark matter halo mass ($M_{200}$), but this prediction has not yet been tested observationally. We present the first observational test of the predicted $R_1$--$M_{200}$ relation and assess whether galaxy size, defined through $R_1$, can serve as a robust estimator of dark matter halo mass. We analyze galaxies from the SPARC database, combining high-quality rotation curves to infer dark matter halo masses with stellar surface density profiles derived from photometric data to measure $R_1$. We then compare the observed relation with the predictions from state-of-the-art cosmological simulations. We find good agreement between observations and simulations: galaxies follow the same tight $R_1$--$M_{200}$ relation predicted by numerical models across the explored mass range, with low intrinsic scatter. We provide the first observational validation of the $R_1$--$M_{200}$ relation, demonstrating that $R_1$ is a simple, physically motivated, and robust estimator of dark matter halo mass. Because $R_1$ is determined entirely from photometric observations, it provides a tool for inferring halo masses from imaging data alone, with applications to studies of galaxy formation and evolution.
\end{abstract}



\section{\bfseries Introduction}

It has long been established that observed galaxies follow a distinct stellar mass-size relation \citep{shen2003,courteau2007}, where size is traditionally defined by the half-mass radius. However, recent studies indicate that this relation becomes significantly tighter when galaxy size is instead defined by the radius at which the stellar surface density drops to $1~\Mopcs$, denoted as $R_1$ \citep{trujillo2020}. The surface density threshold is a proxy of the underlying physical process of star formation. It indicates, for spiral galaxies of the size of the Milky Way, the extent of past or current in-situ star formation on the disc.

The tightness of the relation, with scatter $0.06~\dex$ over more than four orders of magnitude in stellar mass, has also been shown to hold true within hydrodynamical cosmological simulations \citep{arjona-galvez2025,dallavecchia2025} independently of the numerical model for galaxy formation. What drives the relation is the gas surface density threshold that dictates the onset of star formation, as theoretically predicted by \cite{schaye2004}. In addition, \cite{sanchezalmeida2020} showed that, at the threshold of stellar surface density that define $R_1$, S\'ersic profiles \cite{sersic1968} for the same stellar mass and different indices tends to cross, giving similar values of $R_1$. 

Interestingly, analysis of the EAGLE simulations indicates that $R_1$ also closely correlates with the mass of halos hosting galaxies with scatter $\sim 0.1~\dex$ \citep{dallavecchia2025}, significantly less than when using the effective radius \citep{kravtsov2013}. This hints at the tight relation between halo size, angular momentum and gas cooling for galaxy growth. \cite{trujillo2020} and \cite{dallavecchia2025} also showed that, with this definition of size, the relation holds with the same scatter for early-type galaxies, although with different slope.

To date, this predicted relation has not been empirically tested in observed galaxies. In this work, we use rotation curves from the SPARC database \citep{lelli2016}  to derive halo masses, alongside the accompanying photometric maps to derive stellar surface density profiles and determine $R_1$, in order to establish an observational galaxy size--halo mass relation and test it directly against theoretical predictions from simulations.

 We describe the methodology in Section~\ref{sec:methods} and the results in Section~\ref{sec:results}. Discussion and conclusions are in Section~\ref{sec:conclusions}.

\section{\bfseries Methods} \label{sec:methods}
Throughout the work, we define the halo mass as $M_\mathrm{200}$, the mass contained in a spherical region of radius $R_\mathrm{200}$, where the density is 200 times the critical density of the universe. The EAGLE simulations \citep{schaye2015,crain2015}  asssume a $\Lambda$CDM universe with parameters: $\Omega_\mathrm{0}=0.307$, $\Omega_\Lambda=0.693$, $\Omega_\mathrm{b}=0.04825$, $h=0.6777$, $\sigma_8=0.8288$ and $n_\mathrm{s}=0.9611$ \citep{planckxvi}. 

\subsection{\bfseries The SPARC Sample and Stellar Masses}
The Spitzer Photometry and Accurate Rotation Curves (SPARC) database \citep{lelli2016} provides 175 \ion{H}{i} rotation curves obtained from aperture synthesis radio observations spanning over three decades. The sample encompasses a variety of disk galaxies, including dwarf irregulars and high-surface-brightness spirals, spanning broad ranges in stellar mass ($3\times10^{7} \le M_{*}/\mathrm{M}_{\odot} \le 3\times10^{11}$) and rotational velocity ($20 \le V_{\mathrm{rot}}/(\kms) \le 300$).

Rotation curves within SPARC are categorized into three quality flags ($Q$). $Q=1$ represents high-quality, symmetric rotation curves. $Q=2$  represents intermediate quality, typically constrained by lower spatial resolution or minor kinematic asymmetries. $Q=3$ represents low-quality, subject to significant non-circular motions or asymmetric trends. The three quality categories are highlighted throughout our analysis.

SPARC also provides high-quality surface photometry at $3.6\,\mu\text{m}$, which serves as an excellent tracer of the underlying stellar mass distribution, as the stellar mass-to-light ratio ($\Upsilon_*$) remains nearly constant at this wavelength regardless of variations in stellar mass and morphology. The observed $3.6\,\mu\text{m}$ surface-brightness profiles are converted into stellar surface density distributions ($\Sigma_*$), corrected for inclination  (see \citealt{lelli2016}).

\subsection{\bfseries Derivation of the Size Metric $R_1$}

We define the size of each galaxy by the outer radius $R_1$, where the azimuthally averaged stellar surface density reaches $\Sigma_* = 1~\Mopcs $\citep{trujillo2020}. We derive $R_1$ for SPARC galaxies by finding the radius at which the observationally mapped total stellar surface density profile drops to this threshold value.

\subsection{\bfseries Dark Matter Halo Mass Fitting}

To determine the total dark matter host halo mass ($M_\mathrm{200}$) for each galaxy, we implement the rotation curve decomposition procedure outlined by \citet{katz2017}. We set a minimum of 5 data points in the rotation curve for a reliable fit. The total observed rotational velocity, $V_\mathrm{obs}(r)$, is modelled as a quadrature sum of the contributing components to the gravitational potential:
\begin{equation}
V_\mathrm{obs}^2(r) = V_\mathrm{gas}^2(r) + \Upsilon_* V_\mathrm{disk}^2(r) + \Upsilon_\mathrm{bul} V_\mathrm{bul}^2(r) + V_\mathrm{halo}^2(r),
\end{equation}
where $V_\mathrm{gas}(r)$ is the contribution of the cold gas (calculated from \ion{H}{i} radial profiles and corrected for helium), $V_\mathrm{disk}(r)$ and $V_\mathrm{bul}(r)$ are the stellar disk and bulge contributions derived from the $3.6\,\mu\text{m}$ surface brightness profiles, $\Upsilon_*$ is the disk stellar mass-to-light ratio at $3.6~\mu\text{m}$ treated as a free parameter in the fit (see below), $\Upsilon_\mathrm{bul} = 1.4\,\Upsilon_*$ is the bulge mass-to-light ratio following the ratio adopted by \citet{lelli2016}, and $V_\mathrm{halo}(r)$ represents the contribution from the dark matter halo.

We parameterize the dark matter component using two density models:

\begin{enumerate}

\item \textbf{Navarro, Frenk \& White (NFW) profile:} Derived from cosmological $N$-body simulations \citep{navarro1996}, the NFW profile provides a two-parameter description of the spherically averaged dark matter density:
\begin{equation}
\rho_{\mathrm{NFW}}(r) = \frac{\rho_\mathrm{s}}{(r/r_\mathrm{s})(1 + r/r_\mathrm{s})^2}\,,
\end{equation}
where $r_\mathrm{s}$ is the scale radius at which the logarithmic slope of the density profile equals $-2$, and the characteristic density $\rho_\mathrm{s}$ is determined by the halo concentration $c_{200} = R_{200}/r_\mathrm{s}$ \citep[see][for details]{navarro1996}.

\item \textbf{Di Cintio et al.~(DC14) profile:} Formulated from high-resolution hydrodynamical simulations \citep{dicintio2014}, the DC14 profile accounts for the modification of dark matter halos driven by supernova feedback and gas outflows. It follows a generalised $(\alpha,\beta,\gamma)$ density distribution:
\begin{equation}
\rho_{\mathrm{DC14}}(r) = \frac{\rho_\mathrm{s}}{(r/r_\mathrm{s})^\gamma \left[1 + (r/r_\mathrm{s})^\alpha\right]^{(\beta-\gamma)/\alpha}}\,,
\end{equation}
where the inner slope $\gamma$, outer slope $\beta$, and transition sharpness $\alpha$ are not fixed constants but vary as continuous functions of the stellar-to-halo mass ratio $M_*/M_{200}$, calibrated directly from hydrodynamical simulations \citep{dicintio2014}. This allows the profile to smoothly interpolate between a cuspy NFW-like shape at high and low $M_*/M_{200}$ and a cored profile at intermediate mass ratios, where supernova feedback is most effective at redistributing dark matter.

\end{enumerate}

The free parameters of these mass models are the halo virial velocity $V_{200}$ (from which $M_{200}$ is derived), the halo concentration $c_{200}$, and the stellar mass-to-light ratio $\Upsilon_*$. These are fitted to the observed rotation curves via a Markov Chain Monte Carlo (MCMC) approach following \citet{katz2017}, using the \textsc{emcee} ensemble sampler \citep{foreman-mackey2013}. 

We impose a Gaussian prior on $c_{200}$ using the concentration--mass relation of \citet{dutton2014} with a scatter of 0.11\,dex. The stellar mass-to-light ratio $\Upsilon_*$ is given a flat prior over the physically motivated range $[0.3, 0.8]~\mathrm{M_\odot/L_\odot}$, consistent with stellar population synthesis models for a Chabrier IMF \citep{chabrier2003} at $3.6~\mu\text{m}$. Although $\Upsilon_*$ is nearly constant at this wavelength, allowing it to vary within this narrow range self-consistently accounts for galaxy-to-galaxy variations in stellar population age and metallicity. We verify \textit{a posteriori} that the resulting posterior distributions cluster around $\Upsilon_{*,\mathrm{disk}} \approx 0.45~\mathrm{M_\odot/L_\odot}$ and $\Upsilon_{*,\mathrm{bul}} \approx 0.65~\mathrm{M_\odot/L_\odot}$, well within the prior range and broadly consistent with the fixed values of $0.5$ and $0.7~\mathrm{M_\odot/L_\odot}$ adopted by \citet{lelli2016} for disk and bulge respectively. The factor of approximately two spanned by the prior is generally considered acceptable for stellar mass estimates at $3.6~\mu\text{m}$ \citep{lelli2016, meidt2014}.

To isolate the dominant posterior mode and avoid degenerate secondary solutions, we apply the HDBSCAN clustering algorithm \citep{campello2013} to the MCMC chains following \citet{katz2017}, retaining only the highest-probability cluster for parameter estimation. Final parameter estimates and uncertainties are taken as the median and 2.5th--97.5th percentile range of the filtered posterior samples.

We then compare the resulting observational measurements of galaxy size ($R_1$), stellar mass, and halo mass to predictions from simulations.

\section{\bfseries Results}
\label{sec:results}

\subsection{\bfseries Size}

\begin{figure}
\centering
\includegraphics[width=0.75\linewidth]{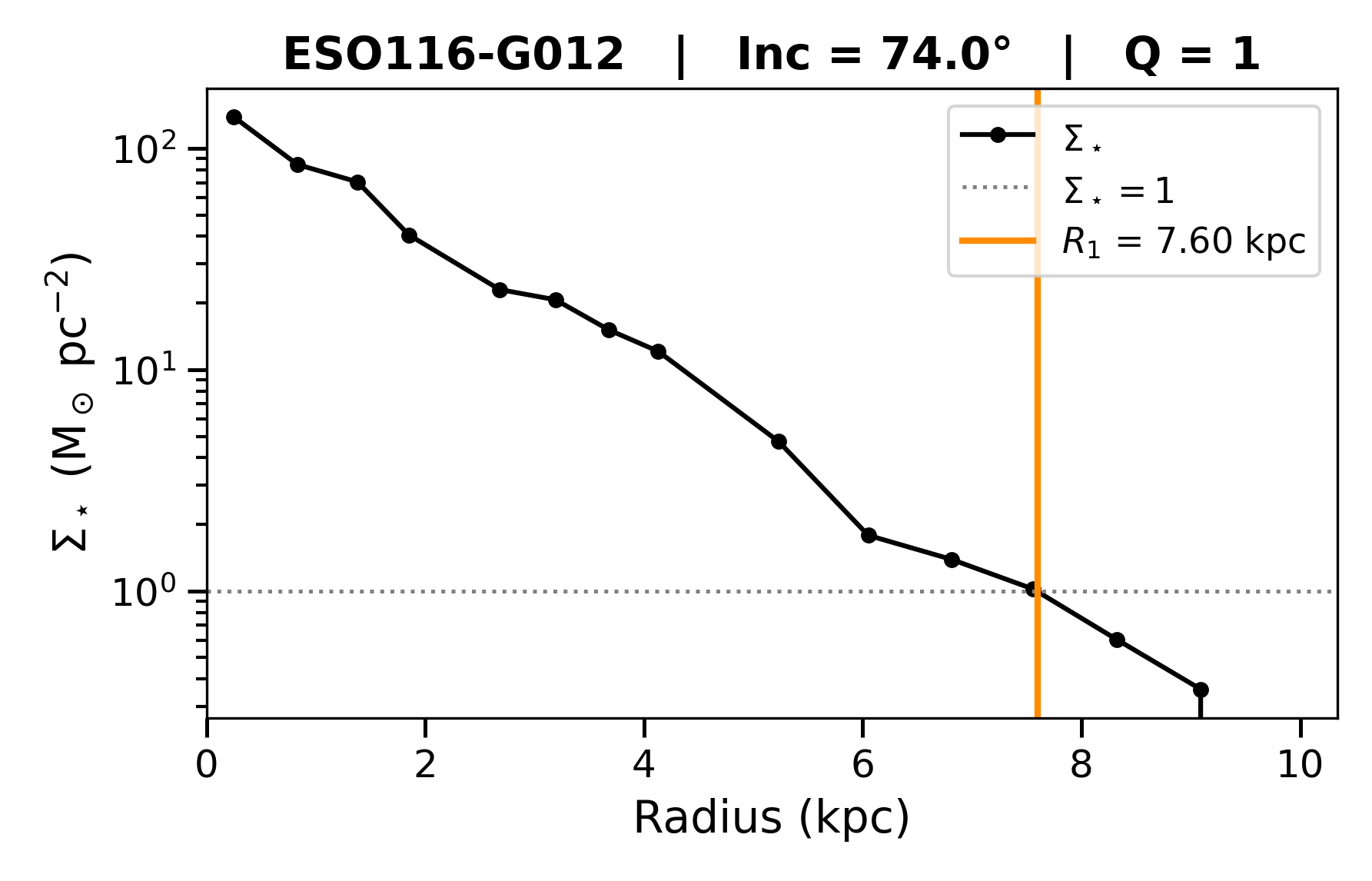}
\caption{\label{fig:f1} \small Stellar surface density profile for ESO116G012, with the dotted horizontal and yellow vertical lines intersecting at $R_1$, the radius at which $\Sigma_\star=1~\Mopcs$}
\end{figure}

In Figure \ref{fig:f1} we show an example derivation for the size, $R_1$, of the galaxy ESO116-G012. In 36 of 175 SPARC galaxies, observations do not go deep enough to reach the threshold surface density for $R_1$. These are discarded, leaving 139 galaxies in our sample.

\begin{figure}[b]
\centering
\includegraphics[width=1.\linewidth]{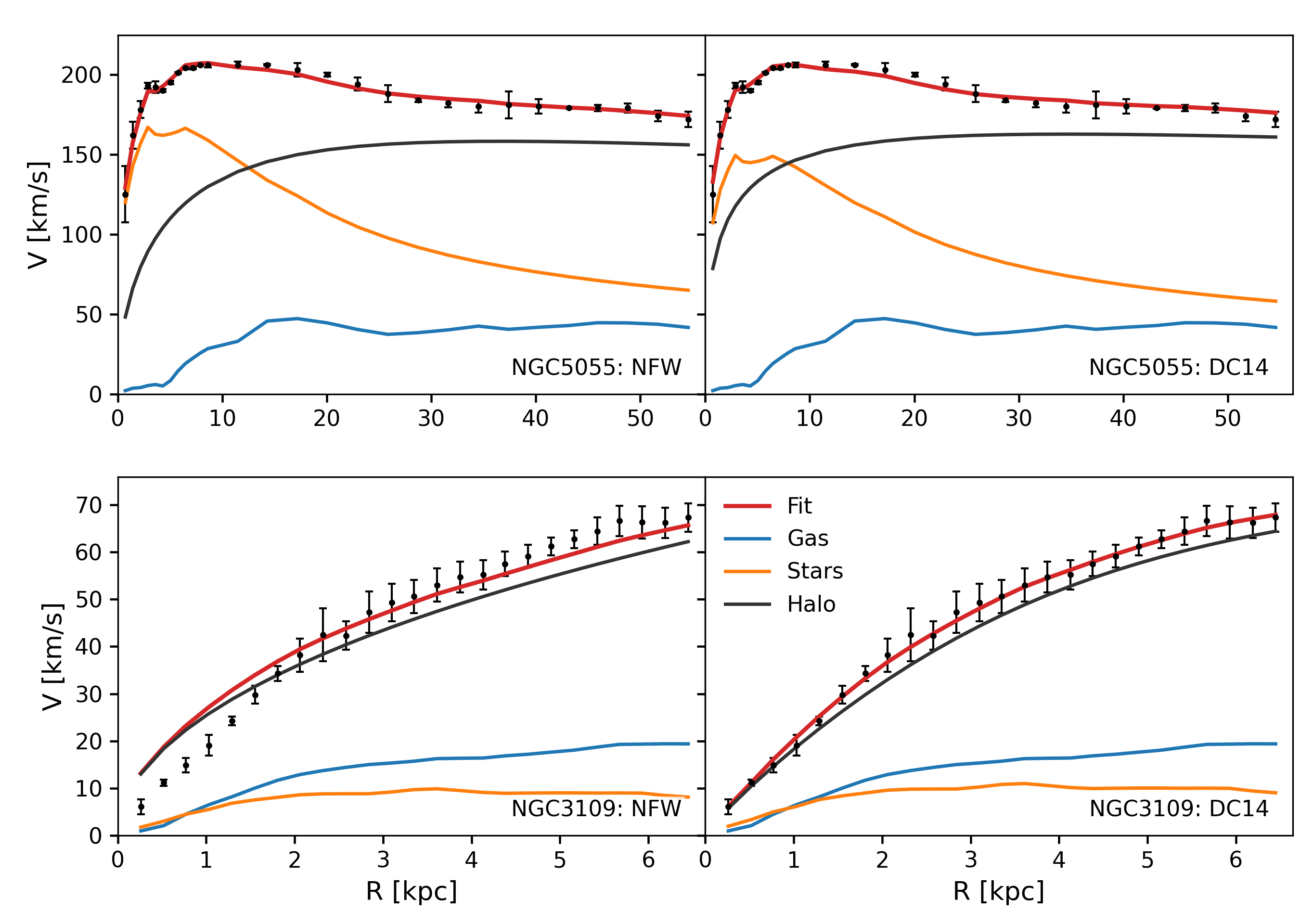}
\caption{\label{fig:f2} Red lines are best-fit NFW (left) and DC14 (right) dark matter halo profiles to rotation curves for a spiral (top) and dwarf galaxy (bottom). The black, orange and blue lines are the contribution to the rotation curve of dark matter, stars and gas mass, respectively. These fits are produced using our implementation of the method of Katz et al. (2017), and we show the same galaxies as in that work to demonstrate the consistency of our results with theirs.}
\end{figure}

\subsection{\bfseries Halo Mass}

Two galaxies (D512-2 and UGC00634) are excluded from the halo mass fitting as their rotation curves contain only 4 data points, leaving 137 galaxies with halo mass determinations. Figure \ref{fig:f2} shows an example of the fits to rotation curves used to derive halo masses, following \cite{katz2017}, as set out in Section~\ref{sec:methods}. Fits using an NFW profile are on the left, whilst fits using the mass dependent DC14 profile are on the right. The spiral galaxy NGC5055 is shown in the upper panel and the dwarf galaxy NGC3109 is shown in the lower panel.

\begin{figure}
\centering
\includegraphics[width=.9\linewidth]{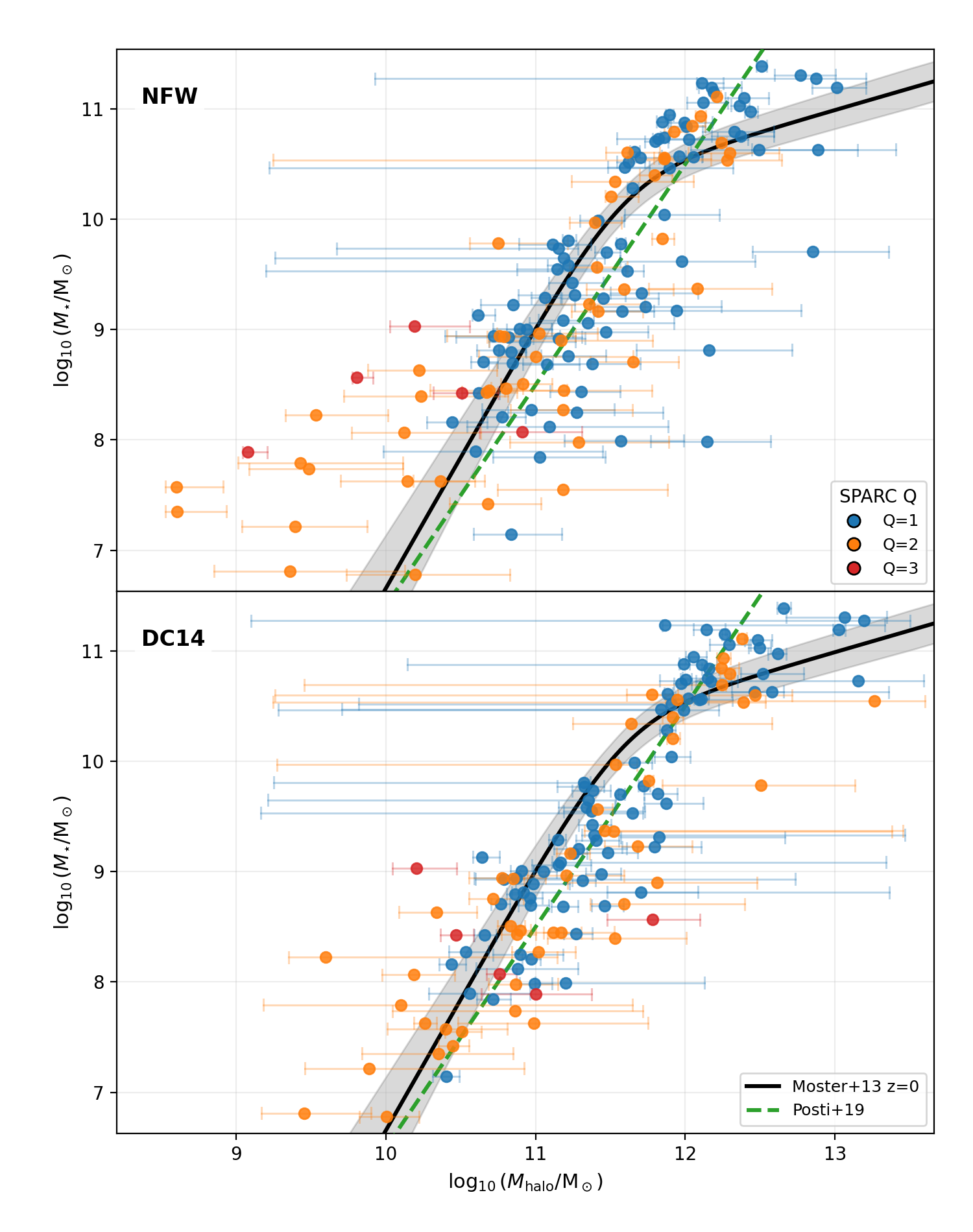}
\caption{\small The relation between our derived $M_{200}$ and stellar mass for all galaxies, using results from NFW fits (top) and DC14 fits (bottom). Colors indicate the quality of rotation curve data from Q1, high quality, Q2 medium and Q3 low quality in blue, orange and red respectively. The \cite{moster2013} relation is shown as a black line, whilst the Posti linear fit for Spiral galaxies is shown as a green dashed line.
\hspace{\textwidth}}
\label{fig:f3} 
\end{figure}

In Figure \ref{fig:f3} we show the relation between the derived $M_\mathrm{200}$ and stellar mass for all SPARC galaxies, using results from NFW fits (top) and DC14 fits (bottom). Colors indicate the quality of rotation curves from SPARC: Q1, high quality, Q2 medium and Q3 low quality in blue, orange and red respectively. Following \cite{katz2017}, which provides full details, error bars represent the 95\% credible intervals ($2.5^{\text{th}}$ to $97.5^{\text{th}}$ percentiles) of the halo mass posterior distribution, determined by isolating the dominant statistical mode of the MCMC chain using HDBSCAN clustering. The black line shows the relation from \cite{moster2013}, whilst the dashed green line shows the power law relation from \cite{posti2019} for spiral galaxies.

\begin{figure}
\centering
\includegraphics[width=.99\linewidth]{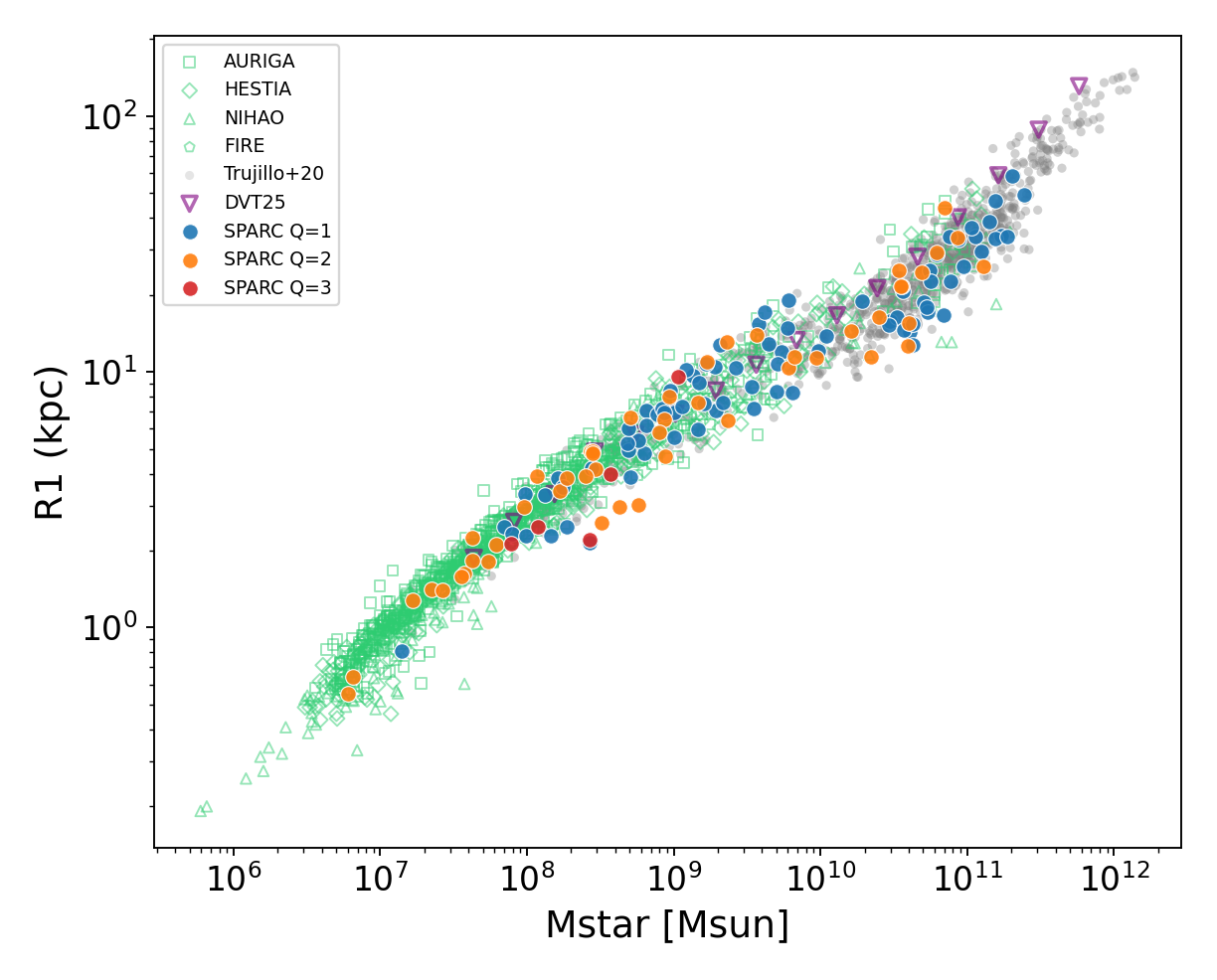}
\caption{\small The relation between $R_1$ and $M_\star$ for SPARC galaxies, overplotted on data from \citealt{arjona-galvez2025}, showing simulation data in green for NIHAO (triangles), FIRE (circle), AURIGA (square) and HESTIA (diamond), data from the EAGLE simulations \citep[purple triangles,][]{dallavecchia2025} and observational results \citep[gray circles,][]{trujillo2020}.} 
\label{fig:f4} 
\end{figure}

\begin{figure}[b]
\centering
\includegraphics[width=.92\linewidth]{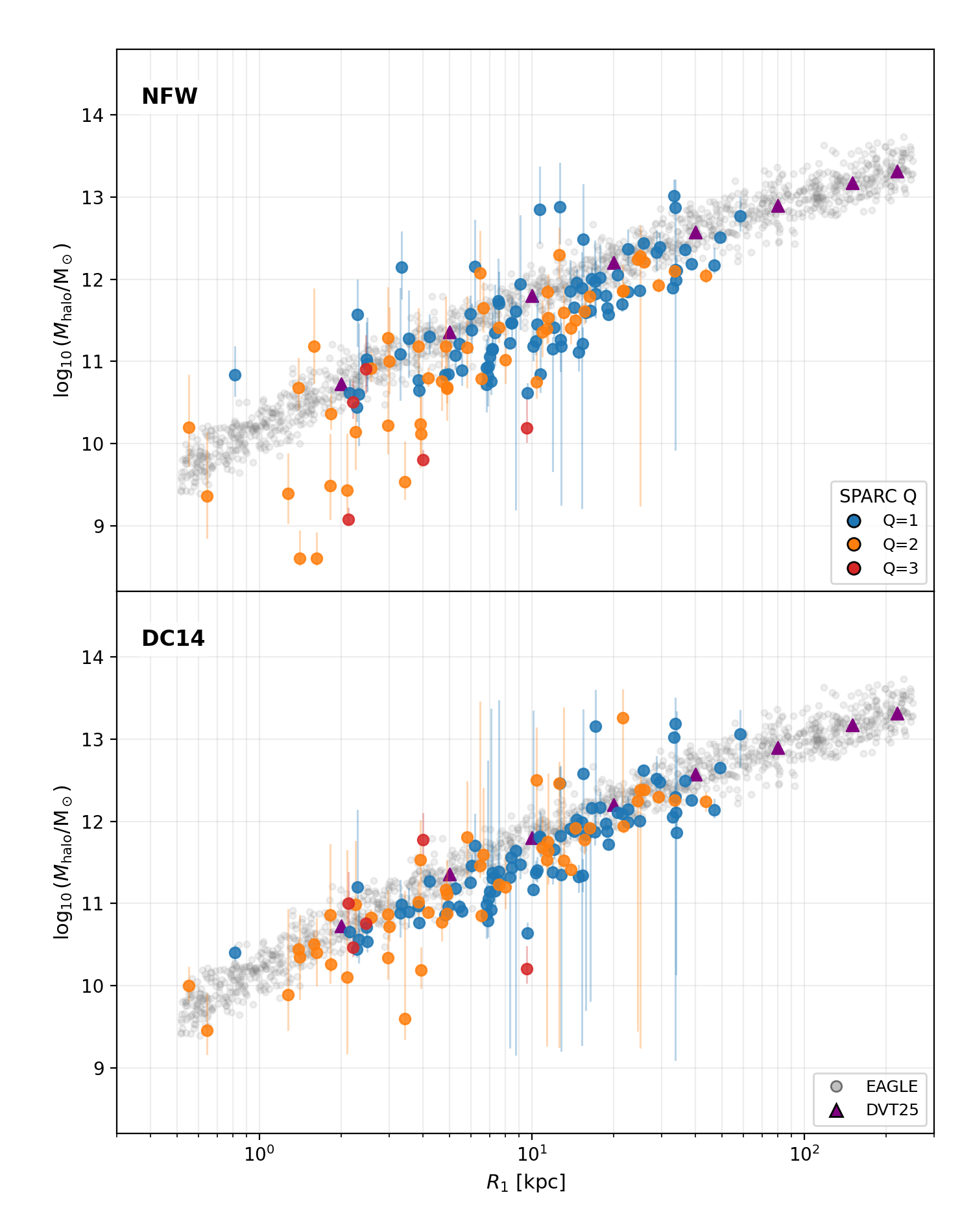}
\caption{\small The relation between $R_1$ and $M_\mathrm{200}$ for SPARC galaxies, overplotted on the relation for EAGLE simulation data (grey circles) and fit to those simulations (purple triangles) both from \cite{dallavecchia2025}.}
\label{fig:f5}
\end{figure}

\subsection{\bfseries $M_\star$--$R_1$ relation}
Figure \ref{fig:f4} shows the relation between $M_\star$ and $R_1$ for the 139 SPARC galaxies that have stellar density profiles that reach as low as $1~\Mopcs$. Also plotted in the background are simulation data in green for NIHAO (triangles), FIRE (circle), AURIGA (square), HESTIA (diamond), all taken from \citealt{arjona2025}, as well as data from the EAGLE simulations \citep[purple triangles,][]{dallavecchia2025} and observational results \citep[gray circles,][]{trujillo2020}. 

\subsection{\bfseries $M_\mathrm{200}$--$R_1$ relation}
Figure \ref{fig:f5} shows the relation between $M_\mathrm{200}$ and $R_1$ for the 137 SPARC galaxies with R1 measurements and halo mass determinations. Also shown is the EAGLE simulation data (gray circles) with binned data as purple triangles, both from \cite{dallavecchia2025}. A very similar relation was found for TNG-50 simulated galaxies \citep{marrero2026}, at least at the high mass end, with a small offset between disc galaxies such as those in the SPARC data set, and bulge dominated galaxies, which have around with around 0.1-0.2 dex smaller $R_1$ at given $M_\mathrm{200}$.

In order to provide a useful prediction that can be used in observations, we provide the best-fit parameters for a double power-law as in \cite{dallavecchia2025}, but inverting the relation shown in that work. We used the following formulation:
\begin{equation}
    \label{eqn:dpl}
    y(R_1) = A\left(\frac{R_1}{R_\ast}\right)^\alpha\left(1 + \left(\frac{R_1}{R_\ast}\right)^s \right)^{(\beta - \alpha)/s},
\end{equation}
where $y(R_1)$ is either $M_{200}(R_1)$ or $R_{200}(R_1)$.
The best-fit parameters are shown in Table~\ref{tab:fitcoeff}. 

\begin{table}
\centering
\caption{\label{tab:eagle_fits} Parameters of the best fits to the EAGLE data for the fit function in Eq.~4.}
\label{tab:fitcoeff}
\smallskip
\begin{tabular}{lcc}
\hline\hline
Parameter & $M_{200}(R_1)$ & $R_{200}(R_1)$ \\
\hline
$A$ & $(4.28\pm0.46)\times10^{11}\,\mathrm{M}_\odot$ & $158.9\pm5.7\,\mathrm{kpc}$ \\
$R_*$ & $16.20\pm0.80\,\mathrm{kpc}$ & $16.20\pm0.80\,\mathrm{kpc}$ \\
$\alpha$ & $1.27\pm0.04$ & $0.42\pm0.01$ \\
$\beta$ & $2.42\pm0.03$ & $0.80\pm0.01$ \\
$s$ & $4.71\pm1.36$ & $4.71\pm1.36$ \\
\hline
\end{tabular}
\end{table}

\subsection{\bfseries V$_\mathrm{max}$--$R_1$ relation}

\begin{figure}
\centering
\includegraphics[width=.9\linewidth]{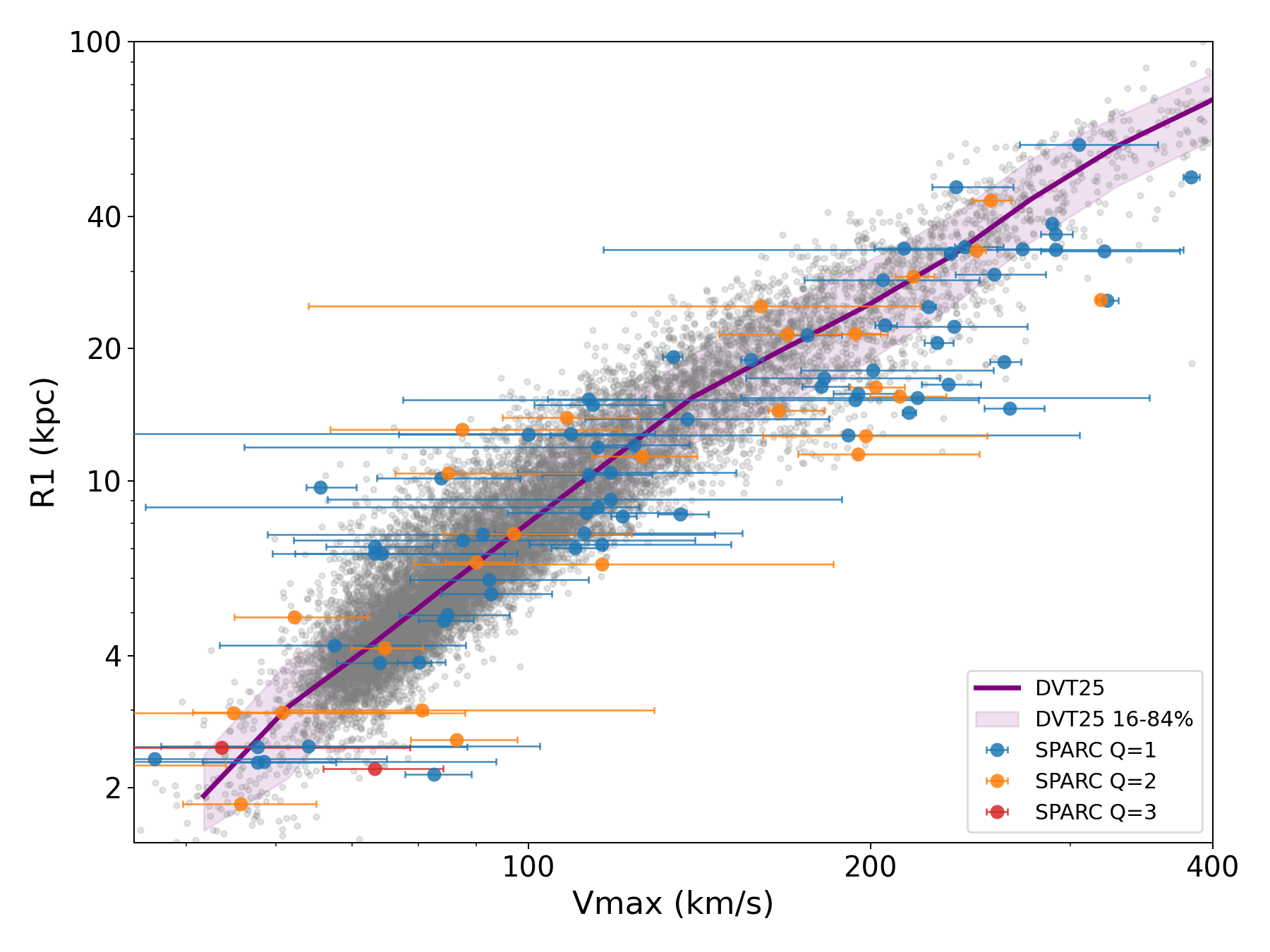}
\caption{\small The relation between $V_\mathrm{max}$ and $R_1$ for SPARC galaxies, plotted over EAGLE simulation data, plotted as grey circles with a non parametric fit shown as a purple line, taken from \cite{dallavecchia2025}. $\qquad$ $\qquad$ $\qquad$
}
\label{fig:f6} 
\end{figure}

Figure \ref{fig:f6} shows the relation between $V_\mathrm{max}$ and $R_1$ for SPARC galaxies, plotted over EAGLE simulation data (grey circles) with a non-parametric fit shown as a purple line, taken from \cite{dallavecchia2025}.

\section{\bfseries Conclusion}
\label{sec:conclusions}
In this work, we have empirically tested the predicted relation between galaxy size ($R_1$, the outer radius where the stellar surface density drops to $1~\Mopcs$) and dark matter halo mass using observed disk galaxies from the SPARC database. With their high-quality 3.6 $\mu\mathrm{m}$ photometry, we determined $R_1$ for 137 galaxies that have surface-density profiles reaching down to the required threshold and rotation curves with 5 or more data points. By decomposing the observed rotation curves of these galaxies via an MCMC framework, we modelled their total host halo masses using both NFW and DC14 dark matter density profiles.

The observational results show that:
\begin{itemize}
    \item The derived $M_\mathrm{200}$ versus stellar mass distribution for SPARC galaxies aligns well with standard semi-empirical and empirical expectations.
    \item The observed SPARC sample follows the tight $M_\ast$--$R_1$ scaling relation predicted by cosmological hydrodynamical simulations and previous observational work.
    \item The observed galaxies exhibit an $R_1$--$M_\mathrm{200}$ relation that is in good agreement with the theoretical predictions and parameterizations derived from the EAGLE simulations.
    \item An alternative, simpler kinematic indicator, the peak rotational velocity $V_\mathrm{max}$, correlates remarkably well with $R_1$, matching simulated trends.
\end{itemize}

Overall, our analysis confirms that the observationally accessible galaxy size metric $R_1$ provides a robust and reliable estimator of both host dark matter halo mass and maximum rotational velocity. This empirical validation establishes $R_1$ as a tool for linking the visible properties of disk galaxies to their underlying dark matter halos.

\vspace{\baselineskip} 
\section*{\bfseries Acknowledgments}
CB and CDV were supported by the Spanish Ministry of Science, Innovation and Universities (MICIU / Agencia Estatal de Investigación, AEI / 10.13039/501100011033) through grant PID2024-156100NB-C22. ADC and EAG acknowledge financial support from MICIU/AEI/10.13039/501100011033 under the Consolidación Investigadora program (grant CNS2023-144669) and grant PID2024-160009NA-I00. This work is also part of grant CEX2025-001609-S, awarded to the Instituto de Astrofísica de Canarias under the Severo Ochoa Centre of Excellence program and funded by MICIU/AEI/10.13039/501100011033. 

The authors used Claude (Anthropic) as an AI assistant to help 
develop and debug analysis code, to assist in drafting and editing 
the Methods section, and for proofreading and copyediting of the 
manuscript text. All scientific content, results, and conclusions 
were verified and are the responsibility of the authors.

\bibliographystyle{aa}

\bibliography{Refs}

\end{document}